\documentclass[aps,prl,twocolumn,showpacs,superscriptaddress]{revtex4}

\usepackage{psfrag}
\usepackage{graphicx}
\usepackage{amsfonts}
\usepackage[figuresright]{rotating}  
\usepackage{amssymb}
\usepackage{amsmath}
\usepackage{subfigure}
\usepackage{multirow}
\usepackage{tabularx}

\def\LSCO{La$_{2-x}$Sr$_x$CuO$_4$}

\def\YBCO{YBa$_2$Cu$_3$O$_{6+y}$}

\def\BSCCO{Bi$_2$Sr$_2$CaCu$_2$O$_{8+\delta}$}
\def\C60{A$_x$C$_{60}$}

\def\HgCu3{HgCa$_2$Cu$_3$O$_{8+y}$}
\def\HgCu4{HgBa$_2$Ca$_3$Cu$_4$O$_{10+y}$}
\def\TlCu{Tl$_2$Ba$_2$CuO$_{6+\delta}$}
\def\TlCu3{Tl$_2$Ba$_2$Ca$_2$Cu$_3$O$_{10+y}$}
\def\TlCu4{Tl$_2$Ba$_2$Ca$_3$Cu$_4$O$_{12+y}$}

\def\BiCu3{Bi$_2$Sr$_2$Ca$_{2}$Cu$_3$O$_y$}

\def\8LSCO{La$_{1.88}$Sr$_{.12}$CuO$_4$}
\def\110LNSCO{La$_{1.5}$Nd$_{0.4}$Sr$_{0.1}$CuO$_{4}$}
\def\stage4LCO{La$_{2}$CuO$_{4+\delta}$}
\def\Y248{YBa$_2$Cu$_4$O$_8$}

\def\NbSe2{NbSe$_2$}
\def\TaSe2{TaSe$_2$}
\def\TiSe2{TiSe$_2$}
\def\NaCoOH2O{Na$_{0.3}$CoO$_{2y}$H$_2$O}
\def\MgB2{MgB${}_2$}

\def\avg#1{\langle#1\rangle}

\def\nn{\nonumber}

\begin{document}

\title{Spin-charge interplay in electronic liquid crystals: fluctuating spin stripe driven by charge nematic}
\author{Kai Sun}
\affiliation{Department of Physics, University of Illinois at Urbana-Champaign, 1110 West Green Street, 
Urbana, IL 61801}
\affiliation{Joint Quantum Institute and Condensed Matter Theory Center, Department of Physics, University of Maryland, College Park, MD 20742}
\author{Michael J. Lawler}
\affiliation{Department of Physics, Applied Physics and Astronomy, Binghamton University, Binghamton, NY 13902}
\affiliation{Department of Physics, Cornell University, Ithaca, NY 14853}
\author{Eun-Ah Kim}
\affiliation{Department of Physics, Cornell University, Ithaca, NY 14853}

\begin{abstract}
We study the interplay between charge and spin ordering in electronic liquid crystalline states with a particular emphasis on fluctuating spin stripe phenomena observed in recent neutron scattering experiments\cite{Hinkov2008, Haug2009}. Based on a phenomenological model, we propose that charge nematic ordering is indeed behind the formation of temperature dependent incommensurate inelastic peaks near wavevector $(\pi,\pi)$ in the dynamic structure factor of \YBCO. We strengthen this claim by providing a compelling fit to the experimental data which cannot be reproduced by a number of other ordering possibilities.
\end{abstract}
\pacs{71.10.Hf,71.45.Gm,75.10.Jm}
\date{\today}
\maketitle 

Kaleidoscopic variety of competing ordering tendencies
 is both a hallmark of  correlated electron fluids, such as cuprate and Fe based superconductors\cite{KivelsonYao2008,Coleman},
 and a theoretical challenge. As such, a clear identification of a broken symmetry phase offers a valuable guiding principle.  Recent observations of temperature, energy and doping dependent onset of anisotropy in 
inelastic neutron scattering (INS) studied by \textcite{Hinkov2008} provides  an opportunity for just such identification. 

\begin{figure}[t]
\begin{center}
\includegraphics[width=0.4\textwidth]{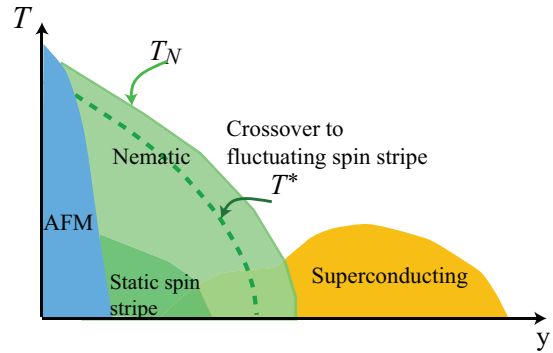}
\end{center}
\caption{(color online) Schematic phase diagram of \YBCO. Here the shaded region bound by $T_N$ represents a nematic phase and the dashed line below $T_N$ represents the crossover temperature $T^*$ to a ``fluctuating spin stripe'' behavior.  
Sufficiently strong nematic ordering at low enough temperatures could further stabilize a static spin stripe phase.
\label{fig:phase_diagram}}
\end{figure}

The symmetry of the ``fluctuating spin stripe'' phenomena (one-dimensional incommensurate spin modulation at finite energy)  observed in Ref. \cite{Hinkov2008} is consistent with that of a nematic phase \cite{Kivelson1998} (a metallic state that  breaks rotational symmetry without breaking translational symmetry).  Furthermore, the qualitative departure in the magnetic response of underdoped \YBCO in Ref. \cite{Hinkov2008} from that of optimally and overdoped regimes $y\gtrsim0.5$ \cite{Hayden:2004p466,Hinkov:2004p2082,Hinkov:2007fk,Tranquada:2004p2083} indicates the possible existence of a quantum critical point at around $y\sim 0.5$ as  we sketch in Fig. \ref{fig:phase_diagram}. 
In specific, low-energy features are enhanced in INS of $y=0.45$ while the extensively studied high energy ``hour-glass'' dispersion which are prominent at higher doping \cite{Hayden:2004p466,Hinkov:2004p2082,Hinkov:2007fk,Tranquada:2004p2083}  is suppressed.
(Considerable attention has been directed towards this resonance feature and its significance. See e.g. Ref.\cite{Tranquada-review}).  
However, despite the reported  temperature dependence of the finite-frequency incommensurability
being suggestive of an order parameter \cite{chain}, it has not been clear how this quantity can be   
related to a specific order parameter since an order parameter is defined by broken symmetry of the ground state.

There has been a number of theoretical studies regarding possible signatures of electronic liquid crystal physics in the magnetic response \cite{yamase:214517,vojta:097001,PhysRevB.72.024502}. While these studies shed light on the hour-glass dispersion observed in $y\gtrsim 0.5$  at high energies, their connection with the low-energy phenomena in the underdoped regime with $y<0.5$ is unclear. Moreover, they focused on the superconducting phase while the observed onset of fluctuating spin stripe behavior is at  $T_{N}\sim 150$K, well above the superconducting ordering temperature $T_c=35$K.

In this letter we propose that charge nematic ordering is the driving force behind the fluctuating spin stripe phenomena observed in underdoped \YBCO. We consider a metallic system proximate to antiferromagnetic (AFM) ordering and show that 
 charge nematic ordering quite uniquely can induce fluctuating and even static spin stripes thus providing a 
   concrete connection between the charge\cite{Ando2002} and spin aspects of liquid crystalline behavior in underdoped \YBCO.
 Our claims are supplemented by a successful fit with available INS data\cite{Hinkov2008}. 

\begin{figure}[h!]
\centering
\includegraphics[width=0.35\textwidth]{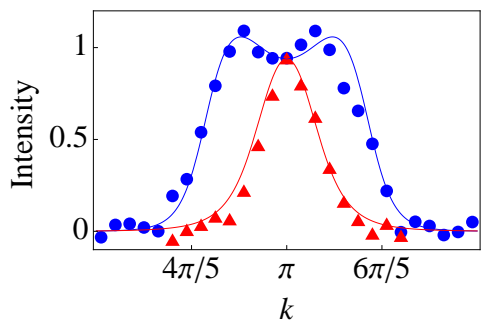}
\caption{(color online) Fit to momentum dependence of INS intensity along $a$ (blue dots) and $b$ (red triangles) axes. Data obtained from Ref.\cite{Hinkov2008} taken at $T=5$K and $\omega=3meV$. Note: any slight asymmetry observed in the data is not accounted for in the phenomenological model presented here.%Momentum dependence of INS intensity along $a$(dots and red line) and $b$(triangles and blue line) axis.  Solid line is our fit to data of Ref.\cite{Hinkov2008} at $T=5$K and $\omega=3meV$.
\label{fig:fit_momentum}}
\end{figure}

{\it Phenomenological Model}\ \ \ 
We start by noting that \textcite{Hinkov2008} detect the dynamic onset of anisotropy through incommensurate inelastic peaks near the AFM wave vector $\mathbf{Q} = (\pi,\pi)$ in a metallic system (see Fig.\ref{fig:fit_momentum}).  Given a microscopic theory of itinerant magnetism being an open question,
 we take a phenomenological approach as a first pass through the problem. In the presence of long-range AFM ordering, low-energy excitation in the particle-hole channel is dominated by gapless spin waves near $\mathbf{Q}$ with infinite life time:
\begin{align}
\vec{\phi}(\mathbf{q},\omega)\!\!=\!\!
\int\!\frac{d^2 \mathbf{k} d\Omega}{2(2\pi)^3}
\psi^\dagger_\alpha(\mathbf{k}\!+\!\mathbf{Q}\!+\!\mathbf{q},\Omega\!+\!\omega)
\vec{\sigma}_{\alpha\beta}
\psi_\beta(\mathbf{k},\Omega).
\label{eq:sdw}
\end{align}
Here the operators $\psi^\dagger_\alpha$($\psi_\alpha$) are the fermion creation (annihilation) operators with spin $\alpha=\uparrow,\downarrow$. $\vec{\phi}$ and $\vec{\sigma}$ are vectors in spin space with $\sigma^i_{\alpha\beta}$ representing $(\alpha,\beta)$ component of Pauli matrix $\sigma^i$ for $i=x,y,z$. 
$\mathbf{q}$ denotes the wavevector of the spinwaves measured from $\mathbf{Q}$ and $\omega$ denotes their frequency.  
For underdoped cuprates outside of but close to
the AFM phase (see Fig.\ref{fig:phase_diagram}), this spin-wave can be either damped or gapped and well defined. 
To quadratic order in $\vec{\phi}$, the appropriate effective action takes the form\cite{disordered-AFM} 
\begin{equation}
S[\vec{\phi}]= \frac{1}{g}\int\frac{d^2 \mathbf{q} d\omega}{(2\pi)^3}
\left(i\Gamma |\omega|+\omega^2-\Delta^2\left(\mathbf{q}\right)\right)
|\vec{\phi}(\mathbf{q},\omega)|^2,
\label{eq:action}
\end{equation}
where $g$ is an overall scale and $\Delta\left({\mathbf{q}}\right)$ is the momentum $\mathbf{q}$ dependent spin-wave gap in the absence of damping and $\Gamma$ is the damping energy scale. Here we take the damping to be of Landau damping form\cite{FL-theory} but most of our conclusions below are insensitive to the details of the damping dynamics. 

The dynamic structure factor measured by INS experiments will be proportional to the spectral function
\begin{align}
\chi''(\omega,\mathbf{q})
=g\frac{\Gamma\omega}{(\Gamma\omega)^2+\left(\omega^2-\Delta (\mathbf{q})^2\right)^2},
\label{eq:spectral_function}
\end{align}
for the action of Eq.~\eqref{eq:action}. The uniform component of the gap $\Delta(\mathbf{q}=0)$ sets the energy scale above which 
INS intensity is significant 
 and the $\mathbf{q}$ dependence of $\Delta(\mathbf{q})$  determine the distribution of intensity in the Fourier space. In the absence of other symmetry breaking tendencies, $\Delta(\mathbf{q})$ will be a minimum at $\mathbf{q}=0$ and respect the point group symmetry of the system. Hence the INS intensity above $\Delta(\mathbf{q}=0)$ will be $C_4$ symmetric (with static orthrhombicity due to chain layer) and peaked at $\mathbf{q}=0$. Temperature and energy dependent anisotropic incommensurability observed in Ref.\cite{Hinkov2008} indicates additional ordering tendencies at play. We first study the effect of charge nematic ordering motivated by  transport anisotropy observed in the nearby regime\cite{Ando2002}. 

In a nematic fluid, charge degrees of freedom collectively break  rotational symmetry of space while preserving the translational symmetry (hence remains metallic) and any other symmetry of the system such as time reversal and spin rotation\cite{Kivelson1998}. A nematic order parameter in a continuum system has the symmetry of charge quadrupole moment ($l=2$ representation of $SO(2)$ rotational group). It has two real components and it can either be written as a symmetric traceless tensor of rank two or as a complex field\cite{Oganesyan2001, Sun2008}.

As the nematic ordering involves spatial symmetry breaking, the nature of the order parameter itself is affected by crystal fields due to the lattice which lowers 
 the spatial rotational symmetry to a discrete group. For instance in a square lattice with  $C_{4v}$  symmetry, the ordering can occur in one of two channels $d_{x^2-y^2}$ or $d_{xy}$ reducing the order parameter symmetry to Ising-like (a single component real field) in either case\cite{kim:184514}. Due to this discretization of the order parameter symmetry, the nematic fluctuation below the transition temperature $T_N$ will be massive.
In \YBCO, the weak external field imposed by the chain layer will likely pick the nematic to occur in the $d_{x^2-y^2}$ channel and a representative form of order parameter can therefore be written as
\begin{equation}
N =\int \frac{d^2\mathbf{k}}{2(2\pi)^2}\bar{N} (\cos k_x-\cos k_y) \psi^\dagger_\alpha(\mathbf{k})  \psi_\alpha(\mathbf{k}).
\label{eq:Nop}
\end{equation}
However, it is important to note that any electronic quantity that is odd under $90^\circ$ spatial rotation such as effective mass anisotropy ratio $(m_y-m_x)/(m_y+m_x)$ or transport anisotropy $(\rho_{xx}-\rho_{yy})/(\rho_{xx}+\rho_{yy})$ can serve as the order parameter for the nematic state\cite{Oganesyan2001}. In fact, short of probes that couple directly to charge quadrupole moments, nematic phases so far has been mostly detected through temperature dependent in-plane transport anisotropy in quasi 2D systems\cite{PhysRevB.65.241313,Borzi:2007fk}.

In \YBCO, \textcite{Ando2002} observed that transport anisotropy increases upon under doping below $y\sim0.5$ while the effect of CuO chains diminishes. Its electronic origin is further supported by the fact that this anisotropy only sets in below a doping dependent onset temperatures $T_N$ ($180$K for $y=0.45$) making nematic ordering a compelling candidate for background broken symmetry in underdoped \YBCO as advanced in Ref.~\cite{Hinkov2008} without a theoretical model. 

Consider the effect of $d_{x^2-y^2}$ nematic ordering below $T_N$ on the spin fluctuations in Fig.~\ref{fig:phase_diagram}.
The effective action now depends on both $\vec{\phi}$ and $N$. Away from the classical critical region near $T_N$, the gapped nematic fluctuations can be integrated out and  the effect of finite $N$ can be represented by $N$-dependent coefficients in the gradient expansion of  $S[\vec{\phi},N]$ in the long distance limit. We represent such $N$ dependence of $S[\vec{\phi},N]$ using the $N$ dependent ``gap'' $\Delta(\mathbf{q};N)$. On symmetry grounds $\Delta(\mathbf{q};N)$ in the long distance limit takes the following form  to the quadratic order in $q$:
%On symmetry grounds, 
\begin{multline}
\!\!\Delta^2(\mathbf{q};N)\!\!=\!\! \Delta_0^2(N) \!+\! c_0^2(N) q^2 \!-\! c_2^2(N)N (q_x^2-q_y^2) \!+\!\cdots
\label{eq:Delta-N}
\end{multline}
where %all functions of $N$ are even and 
all momenta are in units of the lattice constant (i.e. $q_x \equiv k_x a$).
Assuming tetragonal symmetry of underlying lattice in the absence of symmetry breaking field, all the functions $\Delta_0, c_0, c_2$ should be even functions of $N$ since $N \rightarrow -N$ under $90^\circ$ spatial rotation. 
For small $N$, we expand $\Delta(\mathbf{q};N)$ in powers of $N$ and treat $\Delta_0, c_0, c_2$ as independent of $N$ to lowest order in $N$ \cite{footnote_Vojta}. (Note that this approach would be only valid up to the nematic gap scale which we estimate to be about $\sim18$ meV from the onset temperature for transport anisotropy $T_N\approx180K$.)

Notice that the nematic ordering allows for anisotropic flattening of the momentum dependence of $\Delta(\mathbf{q};N)$ in Eq.~\eqref{eq:Delta-N} which will elongate the inelastic $(\pi,\pi)$ peak in the $x$ direction. For $N>N^*$, with $N^* = c_0^2/c_2^2$, this effect will further shift the peaks to incommensurate positions at $(\pi\pm\delta,\pi)$ at low-energy ($\omega<\Delta(\mathbf{q})$ for any $\mathbf{q}$) due to a dip in $\Delta^2(\mathbf{q};N)$ at $\mathbf{q}=(\pm\delta(T),0)$ with
\begin{equation}
\delta(T) \propto \left(N(T)-N^*\right)^{1/2}.
\label{eq:incommensurability}
\end{equation}
Such incommensurability will naturally have temperature dependence resulting from the temperature dependence of $N(T)$. We represent this crossover temperature $T^*$ for onset of such fluctuating spin stripe at $N(T^*)=N^*$ as a dashed line in Fig.~\ref{fig:phase_diagram}. It is noteworthy that while the form of Eq.~\eqref{eq:incommensurability} is reminiscent of an order parameter, there is no singularity in the free energy at $T^*$ (this is a crossover).

At even lower temperatures, large enough $N$ may even stabilize {\it static} spin stripes. While our lowest order in $N$ analysis may not be applicable in its explicit form for large $N$, it predicts a critical value $N_c$ where $\Delta(\pm\delta,0;N_c)=0$. For $N>N_c$ the system will develop static spin stripes. Such static spin stripes were recently observed at $T=2$ K by \textcite{Haug2009} and a similar effect was also reported in \LSCO\cite{Matsuda2008}
 
\begin{figure}[h]
\centering
\includegraphics[width=0.35\textwidth]{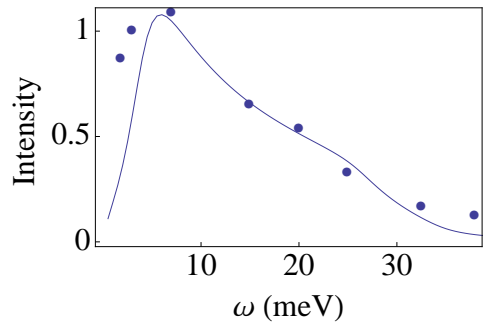}
\caption{(color online) The frequency dependence of the momentum integrated
spectral function.(Note: this is really only valid up to the nematic fluctuation mass scale of $18$ meV).\label{fig:fit_frequency}}
\end{figure}

 {\it Comparison with experiments}\ \ \ 
The model defined by Eqs.~\eqref{eq:action} and \eqref{eq:Delta-N} provides a natural connection between key observations of transport measurements \cite{Ando2002} and recent INS studies in \YBCO \cite{Hinkov2008}.  One non-trivial prediction is the temperature dependence of the incommensurability $\delta(T)$ in the vicinity of $T^*$. 

It is straightforward to show that Eq.~\eqref{eq:incommensurability} implies temperature dependence of $\delta(T)$ reminicsent of that of a mean field order parameter yet without any thermodynamic singularity at $T=T^*$. 
This is independent of the unknown order parameter exponent $\beta$ of $N\propto(T-T_N)^\beta$ and any microscopic details. As long as $N^*$ is small and the spin waves are gapped, so that we may use the quadratic action of Eq.\eqref{eq:action}, the $(N-N^*)^{1/2}$ dependence of $\delta$ in Eq.\eqref{eq:incommensurability} is valid to leading order in $(N-N^*)$. By inserting $N\propto(T-T_N)^\beta$ to Eq.\eqref{eq:incommensurability} , one can show 
%It turns out to behave like a mean field order parameter without a thermodynamic singularity at $T=T^*$. The $(N-N^*)^{1/2}$ dependence of $\delta$ in Eq.\eqref{eq:incommensurability} is valid to leading order in $(N-N^*)$ as long as $N^*$ itself is also small and the spin waves are gapped (so that we may use the quadratic action of Eq.\eqref{eq:action}). Assuming an arbitrary exponent $\beta$ for $N\propto(T-T_N)^\beta$, Eq.\eqref{eq:incommensurability} implies
\begin{equation}
\delta\propto (T-T^*)^{1/2}.
\end{equation}
%This scaling relation is a generic conclusion insensitive to any microscopic details and
%is in good agreement with experimental observation in Ref.\cite{Hinkov2008}.
which is in good agreement with experimental observation in Ref.\cite{Hinkov2008}. 
Thus we have established a model which yields observed $T$-dependent incommensurability.

As a practical test of our phenomenological theory, we fit the INS data of Ref.\cite{Hinkov2008} with the spectral function Eq.\eqref{eq:spectral_function} using $\Delta(\mathbf{q};N)$ from Eq.\eqref{eq:Delta-N}. Fig.\ref{fig:fit_momentum} shows a fit to momentum line cuts of INS data at $\omega=3$ meV and $T=5K$ by setting 
$g/\Gamma = 5.1$ meV, $[\Delta_0^2-(3{\text meV})^2]/\Gamma=2.7$ meV, $c_0^2/\Gamma=16$ meV, $c_2^2N/\Gamma=24$ meV. (We also include symmetry allowed quartic terms of the  form $\lambda_1 (q_x^4+q_y^4) +\lambda_2 q_x^2q_y^2 + \lambda_3 N(q_x^4-q_y^4)$ to fit high $\mathbf{q}$ part of data \cite{footnote_value}.)
Note that this momentum dependence fit at $\omega=3$meV is insensitive
to the relative strength  between damping and gap energy scales
$\Gamma/\Delta_0$. Only the frequency dependence is sensitive to this
ratio.
Similar fits at higher temperatures ($T=40$K and $T=100$K) hint at
thermal fluctuation driven damping at higher temperatures and
$T^*$ being below $100$K (see supplementary online material for details).
Note that we are using a different scheme for extracting $T^*$ from that of Ref.\cite{Hinkov2008}.
 
Fixing most of model parameters by the $\omega=3$meV, $T=5$K data, we fit the frequency dependence of uniform susceptibility $\chi''(\omega)$ at $5$K. 
The energy dependence fit shown in Fig.\ref{fig:fit_frequency} is valid up to the estimated scale of nematic fluctuation mass ($\approx18$ meV). 
Both Figs.\ref{fig:fit_momentum} and \ref{fig:fit_frequency} show good agreement with data. 
However given 
quite a few parameters, we should stress that fitting is more of a check rather than the main result of this paper. Nonetheless, the frequency dependence allows us to estimate $\Gamma/\Delta_0\sim 1.3$ giving a reasonable fit. Note that while one expects the spinwave to be well defined inside superconductor due to lack of low-energy excitations to scatter off, the data of Ref.\cite{Hinkov2008} appears to indicate significant amount of damping. Speculating the source of this damping is beyond the scope of our paper but this might be consistent with existence of the low-energy excitations reported in superconducting underdoped \BSCCO\cite{extinction}.

{\it Effect of Other Ordering Tendencies}\ \ \ 
For completeness, we also consider 
the effect of two other competing orders possibly proximate to the regime of our interest: the charge stripe(smectic) and d-wave
superconductivity.

For charge stripe ordering with wave vector $\mathbf{Q}_{cs}$ the order parameter is 
\begin{align}
\rho_{\mathbf{Q}_{cs}}=
\int\frac{d^2 \mathbf{k} d\Omega}{(2\pi)^3}
\avg{\psi^\dagger_\alpha(\mathbf{k}+\mathbf{Q}_{cs})\psi_\alpha(\mathbf{k})},
\label{eq:CDW}
\end{align}
where the spin index $\alpha$ is summed over. We therefore find an additional contribution to the spinwave action
\begin{align}
\delta S_{cs}=&
-g_{cs}|\rho_{\mathbf{Q}_{cs}}|^2\int\frac{d^2 \mathbf{q} d\omega}{(2\pi)^3}
 |\vec{\phi}(\mathbf{q},\omega)|^2
\nn\\
&-g'_{cs} \rho_{\mathbf{Q}_{cs}} \int\frac{d^2 \mathbf{q} d\omega}{(2\pi)^3} \vec{\phi}(\mathbf{q}-\mathbf{Q}_{cs},\omega) 
\vec{\phi}(-\mathbf{q},-\omega).
\label{eq:CDW_SW}
\end{align}
From Eq.\eqref{eq:CDW_SW} it is clear that the existence of charge stripe shifts the spinwave gap by a constant ($g_{cs}$ term)
and it can also induce fluctuating and static spin stripe with incommensurability fixed by the charge stripe wave vector by $\delta=Q_{cs}/2$ ($g'_{cs}$ term). However, incommensurability so induced will be $T$ independent since it is fixed by $Q_{cs}$. 

In the presence of  $d$-wave pairing $\Delta_d$ the leading additional contribution to the spinwave action is
\begin{align}
\delta S_{sc}=-g_{sc}|\Delta_d|^2\int\frac{d^2 \mathbf{q} d\omega}{(2\pi)^3}
 |\vec{\phi}(\mathbf{q},\omega)|^2.
\end{align}
Hence while $d$-wave superconductivity can also shifts the commensurate spinwave gap $\Delta(\mathbf{q}=0)$ by a constant $g_{sc}|\Delta_d|^2$  and enhance (for $g_{sc}<0$) or suppress (for $g_{sc}>0$) it cannot lead to the observed $T$ dependent incommensurability either. These conclusions can be generalized to other pairing symmetries. 

{\it Discussions}\ \ \ 
We have shown that the existence of charge nematic ordering can take a system proximate to AFM and induce fluctuating spin stripe phenomena: incommensurate INS peaks. Within a symmetry based phenomenological approach we explained the observed $(T^*-T)^{1/2}$ dependence of the incommensurability. Hence, we established a concrete model connecting the spin and charge aspects of the electronic liquid crystalline behavior of \YBCO. We further argued that this is a unique feature of charge nematic ordering.  

One natural implication of our model is the possibility that {\it outside} but close to the charge nematic phase, impurities could stabilize fluctuating or even static spin stripes. Since impurities break the crystal lattice symmetry, they naturally provide a symmetry breaking field which will induce a finite $N$. In weak coupling, this would be given by $N = \chi_N h_N$ where $h_N$ captures the orientational symmetry breaking of the impurities (see Ref.  \onlinecite{Kivelson1998} for a similar argument applied to charge stripes). Since the response $\chi_N$ to this perturbation is expected to be large near a nematic phase, an $N$ larger than $N^*$ could be induced and fluctuating spin stripes will result. If it is further larger than $N_c$, static spin stripes will be induced. 

Several future directions include, establishing a microscopic model which reproduces our phenomenological model in the long-wavelength limit and studying the effects of a magnetic field. This latter question is particularly important in the context of observations of a magnetic field stabilizing static spin stripe phase\cite{Haug2009}. Since such a phase is accessible from our model, it will be interesting to include the effect of magnetic field.  

Furthermore, our analysis may provide a starting point for investigating the interplay between spin and electronic liquid crystalline ordering in other systems.  By now there are a growing number of candidate correlated systems for such interplay, including Mn-doped Sr$_3$Ru$_2$O$_7$\cite{Hossain09} and Fe-based superconductors\cite{Zhao09}.

{\it Acknowledgments} \ \ \ 
We are grateful to E. Fradkin, L. Fritz, V. Hinkov, S. Kivelson, S. Sachdev, J. Tranquada for useful comments and discussions.  
This work was supported in part by the DOE under Contracts DE-FG02-91ER45439 at the University of Illinois and by NSF-PFC at JQI, Maryland (KS) and by the Cornell Center for Materials Research through NSF under Grant No.0520404 and by the KITP through NSF under Grant No. PHY05-51164(EAK and MJL)

\appendix
\section{Supplementary Material}
Here we discuss higher temperature fits to the inelastic neutron
scattering data of Ref.\cite{Hinkov2008}.
Short of explicit temperature dependence in our phenomenological
model, we consider two approaches: (a) we assume implicit temperature dependence of all symmetry allowed phenomenological parameters used in fitting the low temperature ($T=5$K) data; (b) we
assume the dominant effect of temperature is to introduce static damping due to thermal fluctuation. 

\begin{figure}[h]
\begin{center}
\includegraphics[width=0.48\textwidth]{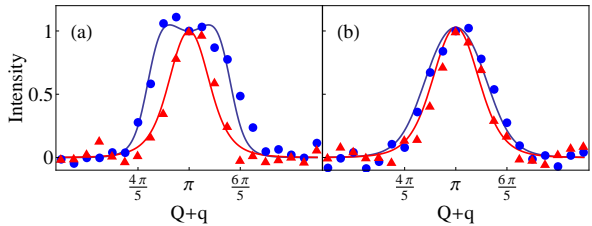}
\end{center}
\caption{ Fit to momentum dependence of INS intensity along $a$ (blue dots) and $b$ (red triangles) axes. Data obtained from Ref.\cite{Hinkov2008} taken at (a) $T=40$K and (b) $T=100$K with $\omega=3$ meV. 
}
\label{fig:fitting1}
\end{figure}

Assuming implicit $T$-
dependence of fitting parameters, we fit the  momentum dependent neutron scattering 
data taken at temperatures
$T=40$K and $100$K. For the fits shown in Fig. \ref{fig:fitting1}, the parameters are
$g/\Gamma = 5.8$ meV ($8.3$ meV), $[\Delta_0^2-(3{\text
  meV})^2]/\Gamma=2.9$ meV ($3.9$ meV), $c_0^2/\Gamma=10$ meV ($15$
meV), $c_2^2N/\Gamma=15$ meV ($5$ meV), $\lambda_1/\Gamma=26$ meV
($14$ meV), and $\lambda_3/\Gamma=18$ meV ($2$ meV) for  $T=40$K ( and
for $T=100$K respectively).

At higher temperatures we observe from this fit that  $c_0>c_2$ at T = 100K
while $c_0<c_2$ for both $40$K and $5$K.
As $c_0<c_2$ is necessary for a double peak structure
separated by $\delta$ (our definition of incommensurability, see Eq. (3) of text),
we interpret the $100$K fit to be above
the crossover temperature scale $T^*$ below which the structure factor
shows a double peak. Assuming $T_N\approx150K$, the system is then in a nematic phase with
anisotropic structure factor at $T=100$K and above the crossover to
the low temperature fluctuating stripe behaviour. 

\begin{figure}[h]
\begin{center}
\includegraphics[width=0.35\textwidth]{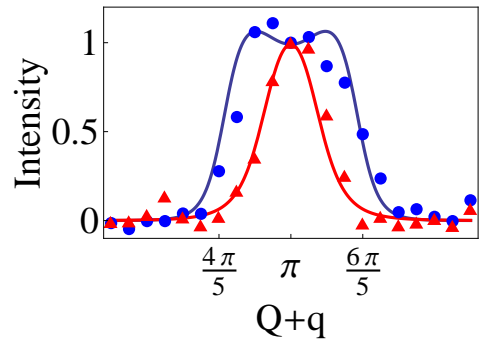}
\end{center}
\caption{Single-parameter fitting to momentum dependence of INS intensity 
along $a$ (blue dots) and $b$ (red triangles) axes at $T=40$K and $\omega=3$ meV with phonon
scatterings. }
\label{fig:fitting2}
\end{figure}

Consider the second approach to fitting mentioned above. Since $T=40{\rm K}>\hbar\omega/k_B$ for the neutron energy transfer of $\omega=3$meV, sizable damping due to thermal fluctuation is
expected at $40$K compared to $T=5$K. Hence, we incorporate thermal fluctuation driven damping into the effective action through  $\Gamma_0(T)$ such that $\Gamma_0(T=5K)=0$:  
\begin{equation*}
S[\vec{\phi}]= \frac{1}{g}\int\frac{d^2 \mathbf{q} d\omega}{(2\pi)^3}
\left(i\Gamma_0+i\Gamma |\omega|+\omega^2-\Delta^2\left(\mathbf{q}\right)\right)
|\vec{\phi}(\mathbf{q},\omega)|^2.
%\label{eq:action1}
\end{equation*}
It turns out both $5$K data and $40$K data can be fit by fixing all
other parameters except overall intensity to their 5K values and setting $\Gamma_0(T=40K)/\Gamma=1.4$ meV (see Fig. \ref{fig:fitting2}).
This signifies enhanced damping as the main difference between these data sets . 

Both the experimental resolution and the phenomenological nature of our
theory make it difficult for a multi parameter fit to be
deterministic. However, the above demonstrates that our theory offers
a sensible fit to the data for multiple line cuts in momentum space and 
the temperature dependence of the neutron scattering structure factor. In particular, these finite temperature fits demonstrate that $T^*$ in the range of
$40K\lesssim T^*\lesssim 100K$ and additional thermal fluctuation at
$40$K compared to $5K$ are consistent with the data..

%\bibliography{fluctuating_stripe_short}

\end{document}